# Correlation of multiple sclerosis (MS) incidence trends with solar and geomagnetic indices: time to revise the method of reporting MS epidemiological data

**Fahimeh Abdollahi [1], Seyed Aidin Sajedi [2]**

1. Department of Internal Medicine, Golestan Hospital, Ahvaz Jundishapur University of Medical Sciences, Ahvaz, Iran

2. Department of Neurology, Golestan Hospital, Ahvaz Jundishapur University of Medical Sciences, Ahvaz, Iran

**Corresponding author:** Seyed Aidin Sajedi  **Email:** dr.sajedy@gmail.com

**Abstract**

**Background:** Recently, we introduced solar related geomagnetic disturbances (GMD) as a potential environmental risk factor for multiple sclerosis (MS). The aim of this study was to test probable correlation between solar activities and GMD with long-term variations of MS incidence.

**Methods:** After a systematic review, we studied the association between alterations in solar wind velocity ($V_{SW}$) and planetary A index ($A_P$, a GMD index) with MS incidence in Tehran and western Greece, during the 23$^{rd}$ solar cycle (1996-2008), by an ecological-correlational study.

**Results:** We found moderate to strong correlations among MS incidence of Tehran with $V_{SW}$ ($r_S=0.665$, $p=0.013$), with one year delay, and also with $A_P$ ($r_S=0.864$, $p=0.001$) with 2 year delay. There were very strong correlations among MS incidence data of Greece with $V_{SW}$ ($r=0.906$, $p<0.001$) and with $A_p$ ($r=0.844$, $p=0.001$), both with one year lag.

**Conclusion:** It is the first time that a hypothesis has introduced an environmental factor that may describe MS incidence alterations; however, it should be reminded that correlation does not mean necessarily the existence of a causal relationship. Important message of these findings for researchers is to provide MS incidence reports with higher resolution for consecutive years, based on the time of disease onset and relapses, not just the time of diagnosis. Then, it would be possible to further investigate the validity of GMD hypothesis or any other probable environmental risk factors.

**Introduction:**

In spite of epidemiological features that all indicate to the key role of environmental risk factors in the etiology of multiple sclerosis (MS), the nature of these risk factors have remained mysterious. Nevertheless, various hypotheses have been introduced; none of them could explain all features of MS and specially its incidence alterations.[1, 2]

Recently, we introduced a hypothesis based on the probable role of solar related *geomagnetic disturbances* (GMD) as a potential environmental risk factor for MS.[3] Indicating to the evidences of biological effects of GMD, probably through resonance phenomenon or direct effects by impacting on brain magnetosomes, we described how GMD have potential ability to elicit essential components of MS pathophysiology such as provoking cellular immunity without the presence of pathologic antigens and disturbing the function of blood brain barrier. Based on these evidences, we framed *GMD hypothesis of MS* and explained in details how it may explain various features of MS.[3] Moreover, by conducting a vast ecological study on global prevalence of the disease, we showed that GMD hypothesis is the best describer of MS prevalence variations.[3] We also discussed how GMD hypothesis may explain the historical trend of MS incidence, by citing to dramatic increase of solar activities and GMD since 1930s.[3]

Nevertheless, as it was the first time that such a hypothesis was proposed, there was not any clue about the probable correlation between solar activities and GMD with long-term variations of MS incidence. It was a crucial issue because supporters of the current hypotheses about MS, such as ultra-violet B related vitamin D deficiency hypothesis and chronic cerebrospinal venous insufficiency (CCSVI) have not provide evidences of such a relationship, up to now. Therefore, we planned to investigate the presence of such correlations.

As understanding the interactions among solar activities and GMD requires a basic knowledge about space-weather and related matters, for biomedical researchers who feel unfamiliar with these concepts we recommend a simplified, concise and open access description about these issues in a special section of our previous publication that is devoted to *basics of space-weather and GMD*.[3]

**Material and Methods:**

Our aim was to study the association between alterations in MS incidence with solar and GMD indices during the previous completed solar cycle (the 23$^{rd}$ solar cycle, 1996-2008).

Data of Solar activities and GMD, included alterations of solar wind velocity ($V_{SW}$, a main related solar factor) and Planetary A index ($A_P$, a geomagnetic activity index), were retrieved from Goddard spaceflight center-space physics data facility.[4] Then, we calculated annual averages of $A_P$ and $V_{SW}$.

In the other hand, we needed reported data of standardized incidence of MS for consecutive years that covered at least 75% of the mentioned period of 1996 to 2008. Therefore, we searched for reports in PubMed with keywords "Multiple Sclerosis" and "Incidence" in title, completed and published from 2005 onwards.

Based on this search strategy, 68 articles were retrieved from PubMed. Among them just two articles fulfilled the inclusion criteria. Accordingly, data of annual age-adjusted MS incidence of both sexes, for our aimed period, were obtained from previously published data of Tehran (1996-2008) by Elhami et al.[5] and western Greece (1996-2006) by Papathanasopoulos et al.[6] with the kind permission of publisher (copyright © 2008 and 2011 Karger Publishers, Basel, Switzerland).

Then, possible lead-lag relationships among mentioned variables were evaluated by means of cross-correlation analysis[7] for lags between -3 to +3 years. By default, cross-correlation function calculates Pearson's correlation coefficient (r), however, features of Tehran data necessitated analyzing with Spearman's correlation method. Therefore, for Tehran data we determined Spearman's correlation coefficient ($r_S$) between lagged variables according to the result of cross-correlation analysis.

Statistical analysis was performed using the SPSS 16.0 (SPSS Inc., Chicago, IL, USA), and p-value < 0.05 considered as significant in analyses.

**Result**

Annual mean of $V_{SW}$ and $A_P$ had identical pattern of changes. Both showed an increasing pattern from 1996-2002, with subtle wax and wane, and a peak at 2003 that was followed by a decreasing trend in $V_{SW}$ and a semi-stable pattern in $A_p$, toward the end of solar cycle (Fig 1).

Elhami et al. had retrieved annual MS incidence through a retrospective study from the data of Iranian MS society, a non-governmental society that registers MS patients, based on non-mandatory referrals by neurologists. According their report, MS incidence of both sexes in Tehran had been stable from 1996 to 1998, then, it increased smoothly to a maximum in 2005, and decreased by 2008 to a level equal to incidence of 1999 (Fig 1).

Papathanasopoulos et al. had calculated MS incidence in western Greece from the patient records of the department of neurology at Patras University Hospital in Rion- Patras, by conducting a retrospective study that nearly included all cases of MS in the area. They determined MS incidence based on definite MS cases and the year of disease diagnosis. According to their study that reported MS incidence in western Greece up to 2006, alterations in incidence trend were not as smooth as Tehran. There was a short period with decreasing slope from 1996 to 1998, and then a gentle ascending slope till 2003 and a remarkable peak at 2004 can be seen that was followed by a decrease in 2005-2006; a pattern that even visually seems to be very similar to changes in $V_{SW}$ and $A_P$, but, with delay (Fig 1).

By using these sets of data and conducting cross-correlation analysis, we found moderate to strong correlations among MS incidence of Tehran with $V_{SW}$ ($r_S=0.665$, 2-tailed $p=0.013$) with one year lag, and with $A_P$ ($r_S=0.864$, 2-tailed $p=0.001$) with 2 year lag. In addition, and very interestingly, we found very strong correlations among MS incidence data of western Greece with $V_{SW}$ ($r=0.906$, 2-tailed $p<0.001$) and with $A_p$ ($r=0.844$, $p=0.001$), both with one year delay (Fig 2).

**Discussion**

Our result confirmed that there have been notable correlations between MS incidence with solar and related GMD in both locations during 23$^{rd}$ solar cycle. Better correlation in western Greece, in comparison to Tehran, may be related to better case ascertainment in Greece. Because of the fact that incidence data of western Greece has been provided based on the evaluation of the records of all hospitals and neurology clinics in the area, while, MS incidence of Tehran has been provided by non-mandatory patient referrals to Iranian MS society, and therefore, was not representative of the whole MS cases in the area .

Moreover, it may be due to the fact that studied area in western Greece has located about 7° closer to the hot line of geomagnetic disturbances, i.e. geomagnetic 60 degree latitude, in the geomagnetic coordinate system (Fig 3). Considering the result of our previous study[3] that showed MS distribution changes are strongly dependent to distance from this line, in a logarithmic manner, even this amount of difference in distance from this line is expected to induce notable impact on MS epidemiology.

Previously, Investigators of both re-analyzed studies could not explain the cause of observed alterations in their reported incidence data, because of the fact that none of the previous MS hypotheses, even the often-cited vitamin D hypothesis, cannot explain the observed increase in their incidence around 2004-2005 and the following remarkable decrease as well. While, as is obvious in the trends of solar activities and GMD (Fig 1), we can hypothesize that increasing trend of MS incidence has been related to accentuation of solar activities and GMD from 1996 to 2002; and the significant peak of MS incidence in Greece (2004) and in Tehran (2005) was the consequences of high solar activities and GMD in 2003. Accordingly, subsequent decrease of MS incidence after 2004-2005 can also be attributed to the following decreasing trend in solar activity and related GMD.

However, observed correlations in these locations can be regarded as a clue in favor of GMD hypothesis, but, certainly these are not enough for judgment and this fact should be evaluated on more locations. Unfortunately, studies that reported MS incidence for consecutive time are very scarce and nearly all of them contain low resolution data, i.e. annual or periodic data, while, high resolution data of solar and geomagnetic indices are available.

Lack of high resolution data limits the ability to interpret the correlational analysis. For example, however, the observed one year delay between alterations in solar and GMD indices with MS incidence in our study can be considered as a clue in favor of probable causal relationship among risk factor and the disease, but it cannot be regarded as a real delay-time for observing the effect of this probable risk factor. For instance, if a set of severe GMD in the last days of a year cause remarkable increase of incidence in the first days of the following year, analysis of annual data would show a one year delay among risk factor and the disease incidence.

Moreover, our findings indicate that we need a substantial revision in definition and report of incidence in MS research. MS incidence is traditionally reported based on the time of diagnosis of the definitely diagnosed cases according to current MS criteria. In Both re-analyzed articles, the time of diagnosis has been used as the time of disease incidence. But, there is usually a significant delay among MS first attack and the time of confirmation of diagnosis that resulted from the need to meet the criteria of "dissemination in time" in the diagnostic criteria of MS.[9] This lag in confirmation of diagnosis varies case by case. According to authors of re-analyzed articles, during 1999-2003, the average of these time delays have been $1.8 \pm 3.6$ years in Greece and $1.6 \pm 2.8$ years in Tehran. This lag due to calculating MS incidences by the time of definite diagnosis greatly affects the detection of causal relationship in MS epidemiological studies. We believe that these mentioned issues are important obstacles which neutralize any effective tries to find the main MS risk factor(s), not only in the case of testing our GMD hypothesis, but also for testing any other potential environmental risk factor of MS.

To overwhelm these problems, the method of reporting MS incidence should be revised in a way that all incidence reports comprise the data of the time of disease onset, as well. Providing data of the exact time of relapses also seem to be useful and maybe more practical due to the awareness and sensitivity of patients to report promptly any new signs or symptoms to their physicians. However, MS is a relatively rare disease among populations, this aim would be possible by designing nationwide or even worldwide online MS attack registries that receive and report high resolution MS attack data to researchers. Then, it would be possible to conduct superposed epoch analyses to investigate actual abilities and validity of any potential risk factor and related hypothesis in practice.

Finally, it should be reminded that detection of correlation doesn't imply causal relationship and the issue of ecological fallacy should not be forgotten. Nevertheless, it is very important to consider the fact that it is the first time in the history of MS[1, 2, 10] that a hypothesis has introduced an environmental factor that not only can explain possible pathophysiology of this disease and features such as relapsing-remitting nature, birth month effect, migration effect and prevalence distribution,[3] but also can describe its incidence trend, at least in some locations.

**Conclusion**

This evaluation showed the ability of GMD hypothesis in explaining MS incidence alterations, and indicated that we need a substantial revision in the method of reporting MS incidence in future epidemiological researches. Important message of these findings that may be regarded as a new roadmap for MS investigators, especially in the areas near or under geomagnetic 60 degree latitude that experience the most GMD, is to provide long-term reports of MS incidence with higher resolution, i.e. weekly or at least monthly data of MS incidence, based on the exact time of disease onset (time of the first disease attack), and not just the time of diagnosis confirmation.

Clarifying this fact especially for testing the GMD hypothesis may be quiet important, not only for better understanding of MS pathophysiology, but also for designing of possible future trials for prevention of relapses by modifying treatments just before increasing exposure to the risk factor; due to the fact that observational stations in the Earth orbit let us to predict significant GMD before occurrence.

**List of abbreviations**

MS: Multiple sclerosis

GMD: Geomagnetic disturbance(s)

$A_P$: Planetary A index

$V_{SW}$: Solar wind velocity

**Acknowledgement**

No source of funding or grant was used. This study has been presented in world congress of neurology (WCN 2013, Vienna, Austria) and the abstract has been published in WCN abstract book in a supplement of journal of neurological sciences. Authors would like to thank Drs. Mohammad Ali Sahraian, Masoud Etemadifar, Davood Khalili, Nastaran Majdinasab, Seyed Ehsan Mohammadianinejad and Mehdi Jalili for their supports.

**Conflict of interest statement**

The authors declare that they have no competing interests.

**References**

1. Ascherio A, Munger KL. Environmental risk factors for multiple sclerosis. Part I: the role of infection. Ann Neurol. 2007; 61:288-99.

2. Ascherio A, Munger KL. Environmental risk factors for multiple sclerosis. Part II: Noninfectious factors. Ann Neurol. 2007; 61:504-13.

3. Sajedi SA, Abdollahi F. Geomagnetic disturbances may be environmental risk factor for multiple sclerosis: an ecological study of 111 locations in 24 countries. BMC Neurol. 2012;12: 100.

4. OMNIWeb interface. Goddard spaceflight center- space physics data facility. Availabale at http://omniweb.gsfc.nasa.gov

5. Elhami SR, Mohammad K, Sahraian MA, Eftekhar H. A 20-year incidence trend (1989-2008) and point prevalence (March 20, 2009) of multiple sclerosis in Tehran, Iran: a population-based study. Neuroepidemiology 2011; 36:141-7.

6. Papathanasopoulos P, Gourzoulidou E, Messinis L, et al. Prevalence and incidence of multiple sclerosis in western Greece: a 23-year survey. Neuroepidemiology 2008;30:167-7.

7. Bloom RM, Buckeridge DL, Cheng KE. Finding Leading Indicators for Disease Outbreaks: Filtering, Cross-correlation, and Caveats. JAMA 2007;14:76-85.

8. US/UK world magnetic chart-epoch 2000. Available at ftp://ftp.ngdc.noaa.gov/geomag/images/wmmcoor.bmp

9. Polman CH, Reingold SC, Banwell B, et al. Diagnostic criteria for multiple sclerosis: 2010 Revisions to the McDonald criteria. Ann Neurol 2011;69:292-302.

10. Murray TJ. Multiple Sclerosis: The History of a Disease. New York : Demos 2005.

# Figures

**Figure 1.** Age-adjusted MS incidence trends of both sexes during 1996-2008 (reconstructed from data of Tehran[5] and western Greece,[6] with the kind permission of publisher, copyright © 2008 and 2011 Karger Publishers, Basel, Switzerland), and annual solar wind velocity and $A_p$ index from OMNIWeb interface (public domain).[4]

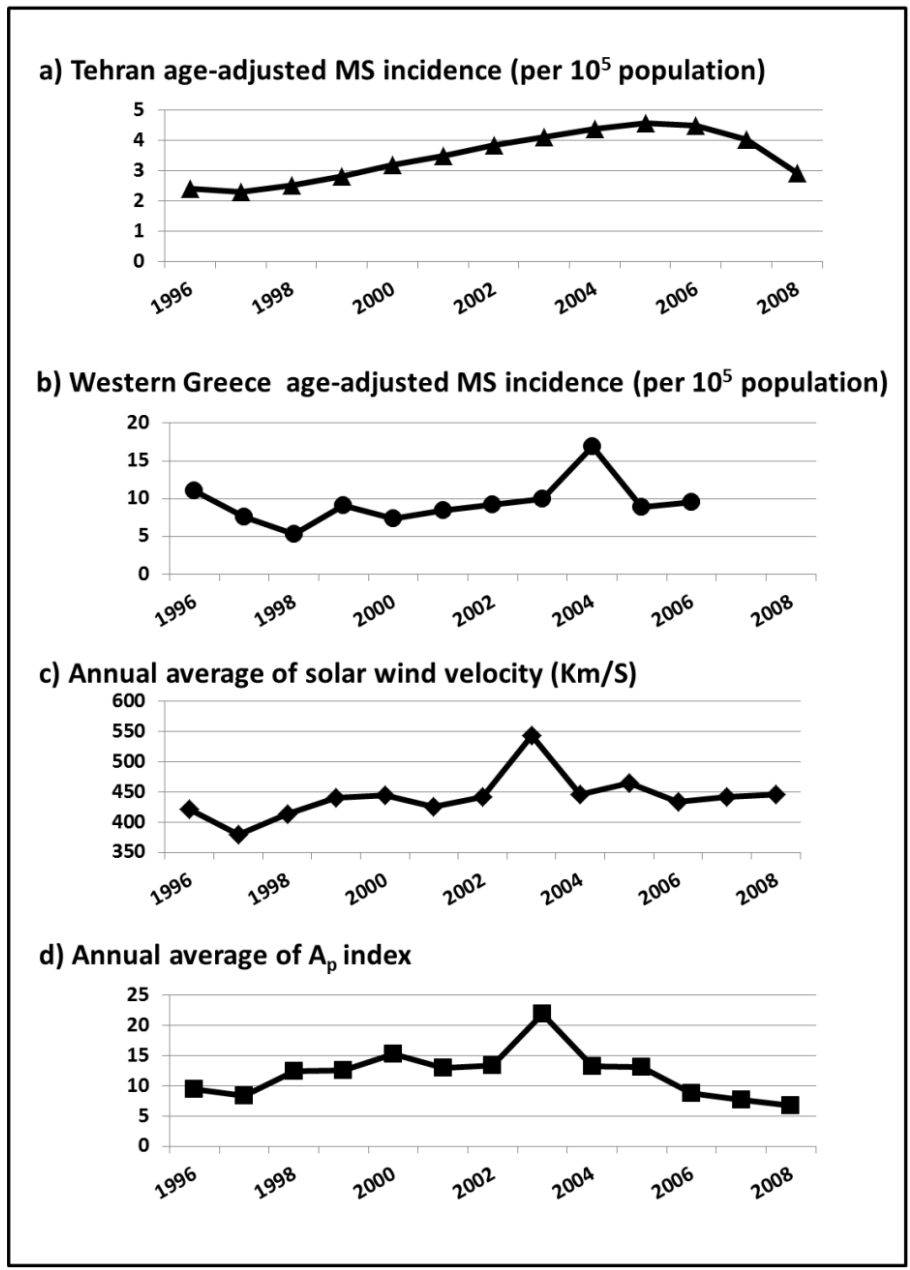

**Figure 2.** Scatter-plots of correlation of both sexes MS incidence data with Solar and geomagnetic indices. Scatter-plots have been constructed after the execution of the mentioned delays on the solar and geomagnetic data sets (please see the text). * As non-parametric Spearman's correlation analysis was used for Tehran data, scatter-plot was constructed by means of the ranked order of mentioned data to show the non-parametric correlation.

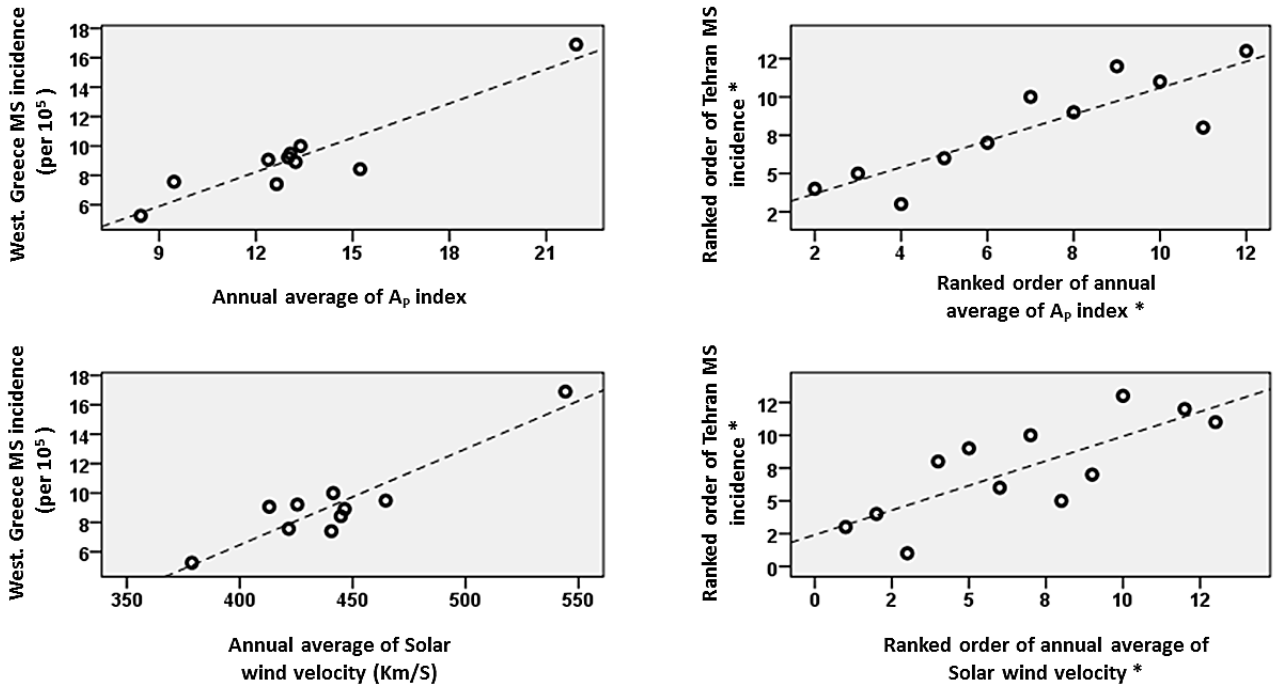

**Figure 3.** Distance of Tehran (white star) and western Greece area (black star) from geomagnetic 60 degree latitude. Note: curve lines indicate geomagnetic latitudes and longitudes. Straight lines illustrate geographic latitudes and longitudes. Reproduced by the kind permission of National Geophysical Data Center.[8]

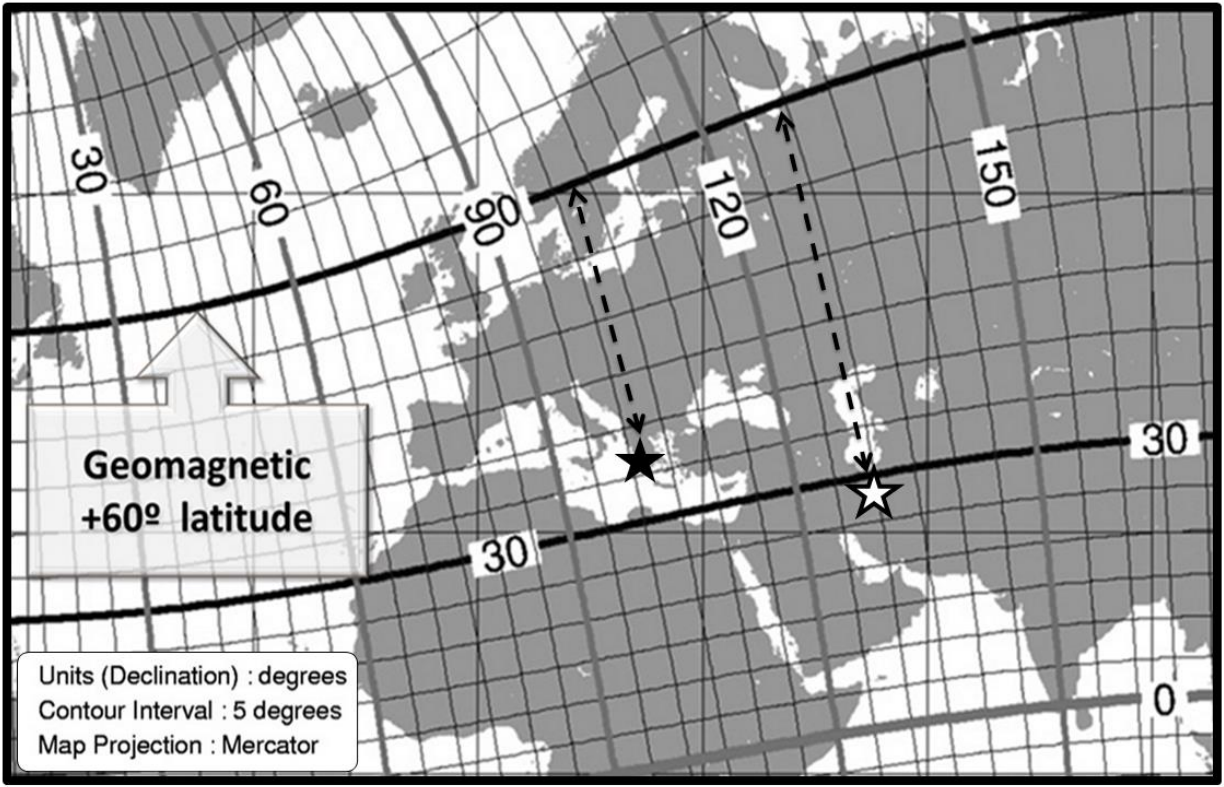